\documentclass[12pt]{article}
\usepackage{amssymb, amsmath}
\usepackage{graphicx}
\usepackage{microtype}
\usepackage{fullpage}
\usepackage{booktabs}
\usepackage{multirow}
\newtheorem{prop}{Proposition}
\usepackage{natbib}
\usepackage{algorithm,algorithmic}
\newcommand{\argmax}{\operatornamewithlimits{arg\,max}}
\newcommand{\argmin}{\operatornamewithlimits{arg\,min}}
\usepackage{url}
\usepackage{xcolor}

\newcommand{\mbx}{\boldsymbol{x}}
\newcommand{\mbX}{\boldsymbol{X}}

\newcommand{\mbB}{\boldsymbol{B}}
\newcommand{\mbP}{\boldsymbol{P}}

\newcommand{\mbD}{\boldsymbol{D}}
\newcommand{\mbT}{\boldsymbol{T}}
\newcommand{\mbb}{\boldsymbol{b}}
\newcommand{\V}{\boldsymbol{V}}
\newcommand{\U}{\boldsymbol{U}}

\newcommand{\mbW}{\boldsymbol{W}}
\newcommand{\R}{\boldsymbol{R}}
\usepackage{setspace}
\newcommand{\mbZ}{\boldsymbol{Z}}{}
\newcommand{\mbv}{\boldsymbol{v}}
\newcommand{\mbu}{\boldsymbol{u}}
\newcommand{\mbA}{\boldsymbol{A}}
\newcommand{\mba}{\boldsymbol{a}}
\renewcommand{\tilde}{\widetilde}
\renewcommand{\hat}{\widehat}

\newtheorem{theorem}{Theorem}

\newcommand{\mbeta}{\boldsymbol{\eta}}
\usepackage{xr}
\title{Dimension reduction for integrative survival analysis}

\author
{Aaron J. Molstad$^{\dagger,*}$ and Rohit K. Patra$^*$ \\
Department of Statistics$^*$ and Genetics Institute$^\dagger$\\
 University of Florida 
}
\date{}

\begin{document}





%
%

\maketitle
\begin{abstract}
We propose a constrained maximum partial likelihood estimator for dimension reduction in integrative (e.g., pan-cancer) survival analysis with high-dimensional predictors. We assume that for each population in the study, the hazard function follows a distinct Cox proportional hazards model. To borrow information across populations, we assume that each of the hazard functions depend only on a small number of linear combinations of the predictors (i.e., ``factors''). We estimate these linear combinations using an algorithm based on ``distance-to-set'' penalties. This allows us to impose both low-rankness and sparsity on the regression coefficient matrix estimator. We derive asymptotic results which reveal that our estimator is more efficient than fitting a separate proportional hazards model for each population. Numerical experiments suggest that our method outperforms competitors under various data generating models. We use our method to perform a pan-cancer survival analysis relating protein expression to survival across 18 distinct cancer types. Our approach identifies six linear combinations, depending on only 20 proteins, which explain survival across the cancer types. Finally, to validate our fitted model, we show that our estimated factors can lead to better prediction than competitors on four external datasets.\bigskip\\
 \textbf{Keywords:}
Cox proportional hazards model, dimension reduction, integrative survival analysis, majorize-minimize, penalty method, reduced-rank regression, variable selection. \\
\end{abstract}

\onehalfspacing
\section{Introduction}
Modeling a patient's survival time based on omic profiles (e.g., transcriptomic or proteomic) is a difficult, yet fundamentally important problem in cancer research. With the advent of personalized medicine, survival models help clinicians plan a course of treatment, and allow patients to make more informed decisions about their care. From a statistical perspective, the challenges are two-fold: first, omic data are often high-dimensional in the sense that the number of predictors (e.g., genes or proteins) can be much larger than the number of patients included in the study (e.g., thousands of genes' expression measured on tens or hundreds of patients). Second, although numerous consortia have begun collecting multi-omic and clinical data on cancer patients (e.g., The Cancer Genome Atlas (TCGA) program), datasets often consist of relatively few patients and have high rates of censoring. Consequently, sample sizes are often too small to identify prognostic predictors or estimate potentially small effects with reasonable accuracy. To address the issues caused by high-dimensionality, small sample sizes, and high censoring rates, one approach is to analyze multiple independent datasets jointly, i.e., to perform an ``integrative'' survival analysis \citep{Liu2014Integrative,zhang2016penalized,maity2020bayesian}. Integrative analyses can improve efficiency and can identify low-dimensional features which are shared across the multiple populations in the study. 

In this article, we propose a new variable selection and dimension reduction method for such integrative survival analyses. Our methodological developments are motivated by a pan-cancer survival analysis in which we model survival as a function of patients' proteomic profiles jointly across 18 distinct cancer types. The proteome contributes to the complex pathophysiology of cancer in ways that cannot be explained by genomic or transcriptomic factors alone \citep{baladandayuthapani2014bayesian}. Recent studies have demonstrated that protein expression levels can be discordant with DNA copy numbers and RNA expression levels \citep{shankavaram2007transcript,akbani2014pan}, so using measurements of protein expression directly may provide novel insights regarding disease progression. More generally, pan-cancer survival models have been of recent interest because they may help identify therapeutic targets shared across cancer types \citep{wang2017multi,maity2020bayesian}.  

To make matters concrete, suppose we are interested in modeling survival in $J$ distinct cancer types. For each $j \in  \{1,\dots, J\}$, let $h_{*(j)}(t \mid \boldsymbol{x})$ be the hazard function for the $j$th cancer type evaluated at time $t$ for a subject with observed protein expression $\boldsymbol{x} \in \mathbb{R}^p$. For any positive integer $m$, we will denote the set $\{1,2,\ldots,m\}$ by $[m]$. We will model $h_{*(j)}(t \mid \boldsymbol{x})$ using the Cox proportional hazards model, i.e., we assume 
\begin{equation} \label{eq:CoxModel}
h_{*(j)}(t\mid \boldsymbol{x}) = h_{*(j)0}(t)  \exp(\boldsymbol{x}^\top\boldsymbol{b}_{*(j)}), \quad j \in [J],
\end{equation}
where $h_{*(j)0}(t)$ and $\mbb_{*(j)} \in \mathbb{R}^p$ are the baseline hazard and regression coefficient vector for the $j$th cancer type, respectively. Let $\boldsymbol{B}_* = (\boldsymbol{b}_{*(1)}, \dots, \boldsymbol{b}_{*(J)}) \in \mathbb{R}^{p \times J}$ be the matrix of unknown regression coefficients. Our proposed method assumes $\mbB_*$ is both low-rank and has many rows entirely zero, i.e., we assume the parametric restrictions
\begin{equation}\label{eq:RR}
{\rm rank}(\boldsymbol{B}_*) \leq r_*  \text{ for } r_* < \min\{p,J\},\quad\text{and}\quad  \|\boldsymbol{B}_*\|_{0,2} \leq s_*  \text{ for } s_* \ll p,
\end{equation}
where for a matrix $\mbA$ with $l$th row $\mbA_{l,\cdot}$, $\|\mbA\|_{0,2} = \sum_{l}\mathbf{1}(\|\mbA_{l,\cdot}\|_2 \neq 0)$ with $\mathbf{1}(\cdot)$ being the indicator function and $\|\mba\|_2$ being the Euclidean norm of the vector $\mba$. That is, for any matrix $\mbA$, $\|\mbA\|_{0,2}$ is the number of nonzero rows of $\mbA$. Under the rank constraint in \eqref{eq:RR}, we can decompose $\boldsymbol{B}_* = \boldsymbol{UV}^\top$ with $\boldsymbol{U} \in \mathbb{R}^{p \times r_*}$ and $\boldsymbol{V} \in \mathbb{R}^{J \times r_*}$, so that we can write the linear predictor for a subject with the $j$th cancer type and protein expression $\boldsymbol{x} \in \mathbb{R}^p$ as $$\boldsymbol{x}^\top\mbb_{*(j)} = (\boldsymbol{x}^\top\mbu_1)v_{(j)1} + (\boldsymbol{x}^\top\mbu_2)v_{(j)2} + \dots + (\boldsymbol{x}^\top\mbu_{r_*}) v_{(j)r_*}, \quad j \in [J],$$
where $\mbu_l \in \mathbb{R}^p$ is the $l$th column of $\U$ and $(v_{(j)1}, v_{(j)2}, \dots, v_{(j)r_*})^\top \in \mathbb{R}^{r_*}$ is the $j$th row of $\V$. We may interpret the $\mathsf{f}_k(\mbx) = \mbx^\top\mbu_k \in \mathbb{R}$ for $k \in [r_*]$ as an unobservable low-dimensional set of ``factors'' which contain all the useful information from $\mbx$ about survival in all $J$ cancer types. In turn, we may then interpret the rows of $\V$ as the distinct regression coefficients for each of the cancer types in the space of the $\mathsf{f}_k$. The $(0,2)$-norm assumption on $\mbB_*$ (i.e., $\|\mbB_*\|_{0,2} \leq s_*$) implies that the same $s_*$ elements of the vectors $\mbu_k$ can be nonzero. Letting $\mathcal{S}_* = \{l: \mbB_{*l,\cdot} \neq 0, l \in [p]\}$, this implies that the factors $\mathsf{f}_k$ depend only on the predictors indexed by $\mathcal{S}_*$, or stated in terms of the hazards, $h_{*(j)}(t \mid \mbx) = h_{*(j)}(t \mid \mbx_{\mathcal{S}_*})$ for all $j \in [J]$, where for any vector $\mba$ and set $\mathcal{S} \subseteq [p]$, $\mba_{\mathcal{S}}$ denotes the subvector of $\mba$ containing only the elements indexed by $\mathcal{S}$.  Together, these assumptions improve parsimony and interpretability. For example, the $k$th factor's biological relevance can be interpreted through the coefficients $\mbu_k$ for $k \in [r_*]$.

Our analysis of the motivating data in Section \ref{sec:DataAnalysis} provides strong evidence that \eqref{eq:RR} is justifiable in the context of pan-cancer proteomics and survival. In particular, cross-validation estimates ($r_*, s_*$) to be $(6, 20)$ (when $J = 18$ and $p=210$). Many of these 20 selected proteins have been identified in the recent literature. Furthermore, examining the estimated factors (e.g. see Figure \ref{fig:factors}) reveals that they contain proteomic information which can distinguish cancer types  (i.e., the factors do not contradict well-established molecular heterogeneity). 

In the next section, we introduce a constrained maximum partial likelihood estimator for $\mbB_*$ under the parametric restriction \eqref{eq:RR} in high-dimensional settings. In subsequent sections, we propose a new algorithm to compute our estimator and establish asymptotic theory thereof. Our work provides contributions on multiple fronts.
 Methodologically, we propose a new framework for jointly modeling survival in distinct populations with high-dimensional predictors. In contrast to related methods, our framework allows practitioners to identify interpretable low-dimensional features shared across populations. 
  Computationally, we provide a new procedure for fitting sparse reduced-rank regression models under nonconvex $L_0$-type constraints. Compared to existing algorithms for related problems, our algorithm has closed-form updates and is conceptually simple as it is based on the majorize-minimize principle \citep[Chapter 1]{lange2016mm}.
   Theoretically, we find the asymptotic distribution of our estimator and quantify the gain in information from exploiting the rank constraint in \eqref{eq:RR}. 
   In contrast to classical results in reduced-rank regression \citep{anderson1999asymptotic}, we do not have a closed-form for the constrained maximum likelihood estimator of $\mbB_*$, so our proof technique is more general.

For the remainder, let $n_{(j)}$ denote the observed sample size for the $j$th population (e.g., $j$th cancer type), and let $t_{(j)1}, \dots, t_{(j)n_{(j)}}$ denote the possibly unobserved survival times for the $j$th population. 
For each $j \in [J]$, we observe $(y_{(j)1}, \delta_{(j)1}, \mbx_{(j)1}), \dots,$ $(y_{(j)n_{(j)}}, \delta_{(j)n_{(j)}}, \mbx_{(j)n_{(j)}})$ 
where $y_{(j)i} = \min(t_{(j)i}, c_{(j)i})$, $c_{(j)i}$ is the censoring time for the $i$th subject in the $j$th population, 
$\delta_{(j)i} = \mathbf{1}(y_{(j)i} = t_{(j)i})$, and $\mbx_{(j)i} \in \mathbb{R}^p$ is the vector of predictors (e.g., protein expression) measured on the $i$th subject from the $j$th population. Let $\mathcal{R}_{(j)i} = \{k: y_{(j)k} \geq y_{(j)i}\}$ be $j$th population's risk set at time $y_{(j)i}$. Finally, let $\|\mbA\|_F^2 = {\rm tr}(\mbA^\top\mbA)$ be the squared Frobenius norm of a matrix $\mbA$.  Throughout, we use upper-case bold letters to denote matrices and lower-case bold symbols to denote vectors. 

\section{Methodology}
\subsection{Estimation criterion}
To fit the proportional hazards models \eqref{eq:CoxModel} under the assumption \eqref{eq:RR}, we maximize a constrained partial likelihood. Specifically, to allow for tied events, we propose to maximize a penalized and constrained version of the Breslow approximation of the partial likelihood. That is, the partial log-likelihood (approximation) we use is $$ \boldsymbol{\mathcal{L}}(\mbB) =  \sum_{j=1}^J \sum_{i=1}^{n_{(j)}} \delta_{(j)i}\bigg[\mbx_{(j)i}^\top\mbb_{(j)} -  \log \Big\{ \sum_{k \in \mathcal{R}_{(j)i}} {\rm exp}\left( \mbx_{(j)k}^\top\mbb_{(j)} \right) \Big\} \bigg].$$ We thus propose to estimate $\boldsymbol{B}_*$ using
\begin{equation}\label{eq:estimator}
\argmin_{\mbB \in \mathcal{C}_r \cap \mathcal{A}_s}\left\{  -\boldsymbol{\mathcal{L}}(\boldsymbol{B}) + \mu \|\boldsymbol{B}\|_F^2  \right\}
\end{equation}
where 
$\mathcal{C}_r = \{\boldsymbol{B} \in \mathbb{R}^{p \times J}: {\rm rank}(\boldsymbol{B}) \leq r\}, \quad  \mathcal{A}_s = \{\boldsymbol{B} \in \mathbb{R}^{p \times J}: \|\boldsymbol{B}\|_{0,2} \leq s \},$
and $\mu > 0$ is a small positive tuning parameter. The ridge penalty, controlled by $\mu$, is used to impose a small degree of shrinkage so that we may consider an effective number of parameters larger than the sample size. In both our simulations and real data analysis, we simply fix $\mu$ to be some small constant (rather than selecting it by cross-validation). The tuning parameters $r$ and $s$, however, should be selected by cross-validation or an information criterion. In Web Appendix B, we propose using a version of the cross-validation criterion from \citet{dai2019cross} to select tuning parameters. 

Though the feasible set $\mathcal{C}_r \cap \mathcal{A}_s$ in \eqref{eq:estimator} is nonconvex, we will later demonstrate that replacing it with a convex approximation leads to an estimator which is often very biased, difficult to tune, and arguably no easier to compute than \eqref{eq:estimator}. For more details, see Web Appendix A.1 and our comparison to such an estimator in Section \ref{sec:SimulationStudies}.

\vspace{-10pt}
\subsection{Related methods and analyses}\label{subsec:related_section}
Dimension reduction methods for survival analysis have primarily focused on single population analyses, e.g., using sufficient dimension reduction \citep{li1999dimension,li2004dimension} or reduced-rank regression \citep{perperoglou2006reduced,fiocco2005reduced}. 
\citet{perperoglou2006reduced} used reduced-rank regression to estimate regression coefficients under the assumption of nonproportional hazards in a single population survival analysis. \citet{fiocco2005reduced} assumed a low-rank decomposition of the regression coefficient matrix under a Cox proportional hazards model for competing risks. \citet{fiocco2005reduced} require that multiple event times are measured on each subject from a single population, whereas we are focused on integrative survival analyses.

Numerous methods exist for the integration of multiple cancer datasets. Many of these methods focus on modeling continuous (uncensored) outcomes and the identification of shared nonzero regression coefficients \citep{zhao2015integrative, Huang2017Promoting}. Similar approaches for variable selection have been proposed under an accelerated failure time (AFT) model \citep{Liu2014Integrative,zhang2016penalized}. For example, \citet{maity2020bayesian} developed a hierarchical Bayesian AFT model which performs variable selection and borrows information across populations through the correlation structure of the prior distributions. Extending these approaches to the Cox model is nontrivial owing to the computational and theoretical challenges of working with the constrained partial likelihood.  Along these lines, \citet{tang2019fusion} proposed a method for fitting a Cox proportional hazards model when there exist predefined subgroups of subjects in a study. Their approach assumes that all subgroups have the same baseline hazard function and assumes that some regression coefficients are equivalent across pairs of subgroups. These assumptions are somewhat restrictive, but lead to an optimization problem that can be solved using existing algorithms and software. 

The work most closely related to our own is the method proposed by \citet{wang2017multi}, who make a low-rank assumption on $\mbB_*$ and propose a nuclear norm penalized maximum partial likelihood estimator. This estimator can impose low-rankness and is the solution to a convex optimization problem, but does not perform variable selection, and thus can perform poorly in high-dimensional settings. Moreover, the nuclear norm penalty imposes global shrinkage towards the origin, so although one can obtain a low-rank estimate of $\mbB_*$, this often comes at the cost of excess shrinkage. We discuss the method of \citet{wang2017multi} and propose a sparse variant thereof in Web Appendix A.1. As we will show in our simulation studies, this approach imposes substantial bias, has poor variable selection performance, and is too difficult to tune to be useful in practice.

Identifying shared prognostic factors across cancer types has also been of recent interest. For example, \citet{hieronymus2018tumor} found that tumor copy number alteration (CNA) burden was a significant prognostic factor in five distinct cancer types. For each $j \in [J]$, \citet{hieronymus2018tumor} effectively assumed 
$ h_{*(j)}(t \mid c_{(j)i}) = h_{*(j)0}(t) {\rm exp}(c_{(j)i} v_{(j)})$
where $c_{(j)i} \in \mathbb{R}$ is the CNA burden for the $i$th subject with the $j$th cancer type and $v_{(j)} \in \mathbb{R}$ is an unknown regression coefficient for the $j$th cancer type.  However, CNA burden is simply the proportion of the genome affected by CNAs, so we can express
$c_{(j)i} = \mbx_{(j)i}^\top 1_p/p$
 where $\mbx_{(j)i} \in \{0, 1\}^p$ is a vector whose $k$th component equals one if there is a CNA at the $k$th genomic location and zero otherwise for each $k \in [p]$, and $1_p \in \mathbb{R}^{p}$ is the $p$-dimensional vector of ones. If instead, we replaced $1_p/p$ with $\mbu \in \mathbb{R}^p$, their hazard function for the $j$th cancer type would be $h_{*(j)0}(t) {\rm exp}\{(\mbx_{(j)i}^\top \mbu) v_{(j)}\}$,
which corresponds exactly to the rank one version of our model, i.e., $\mbB_* = \mbu \mbv^\top$ where $\mbv = (v_{(1)}, \dots, v_{(J)})^\top \in \mathbb{R}^J$.  Thus with \eqref{eq:estimator}, a practitioner could instead estimate both $\mbu$ and $\mbv$ simultaneously, which would allow for a more flexible factor than CNA burden to be discovered. For example, our method could identify genomic locations at which the CNAs are relevant to survival in multiple cancers by taking a (sparse) weighted sum of CNAs rather than simply taking the proportion of the genome affected by CNAs.  In a separate study, \citet{hong2020pan} found that genes among the CASP family serve as useful prognostic factors for breast cancer, hepatocellular carcinoma, and pancreatic cancer. Under \eqref{eq:RR}, this would correspond to taking the $\mbx_{(j)i}$ to be gene expression profiles and assuming $r_* = 1$ with $\mbB_* = \mbu \mbv^\top$, but constraining $\mbu$ to be a vector which has zeros in all positions except those corresponding to genes in the CASP family. Hence, \eqref{eq:estimator} could also discover this type of prognostic factor, but without requiring prior specification of a particular gene family.

Finally, we note that the assumptions in \eqref{eq:RR} are similar to those made in sparse reduced-rank multivariate response regression \citep{chen2012sparse,she2017selective}. The methods of \citet{chen2012sparse} and \citet{she2017selective} are designed for settings where multiple (uncensored) responses are measured on each subject from a single population. As such, these methods cannot be applied in the context of integrative survival analysis. We further elaborate on the distinction between our method and existing methods for sparse reduced-rank regression in Web Appendix H.

\section{Computation}
\subsection{Penalty method based on ``distance-to-set'' penalties}
To compute \eqref{eq:estimator}, we use the penalty method \citep[Chapter 17]{nocedal2006numerical} in concert with the majorize-minimize principle \citep[Chapter 1]{lange2016mm}. The penalty method is especially appealing for \eqref{eq:estimator} since we can employ ``distance-to-set'' penalties, which can be majorized by smooth functions \citep{xu2017generalized,keys2019proximal}. In particular, to compute \eqref{eq:estimator}, we solve a sequence of problems of the form
\begin{equation}\label{penalty_optimProb}
\tilde{\mbB}_\rho =  \argmin_{\mbB\in \mathbb{R}^{p \times J}} \left\{ - \boldsymbol{\mathcal{L}}(\mbB) + \mu \|\mbB\|_F^2 + \frac{\rho}{2} {\rm dist}(\mbB, \mathcal{C}_r)^2 + \frac{\rho}{2} {\rm dist}(\mbB, \mathcal{A}_s)^2\right\},
\end{equation}
where ${\rm dist}(\mbB, \mathcal{C}_r)$ is the Euclidean distance from $\mbB$ and its nearest point in the set $\mathcal{C}_r,$ i.e., 
$$ {\rm dist}(\mbB, \mathcal{C}_r) =\inf_{\mbA \in \mathcal{C}_r}\|\mbB- \mbA\|_F =  \|\mbB - \mathbf{P}_{\mathcal{C}_r}(\mbB)\|_F,$$
with $\mathbf{P}_{\mathcal{C}_r}$ denoting the projection onto $\mathcal{C}_r$. By taking the penalty parameter $\rho \to \infty$, it will occur that $\tilde{\mbB}_\rho \in \mathcal{C}_r \cap \mathcal{A}_s$ as any iterate outside the set $\mathcal{C}_r \cap \mathcal{A}_s$ will lead to a large objective function value. To use this approach in practice, we iteratively compute $\tilde{\mbB}_\rho$ for an increasing sequence of $\rho$ values -- initializing the algorithm for $\tilde{\mbB}_\rho$ at the solution for the previous (smaller) value of $\rho$. In the next subsection, Section \ref{subsec:MMAlg}, we propose a majorize-minimize algorithm for \eqref{penalty_optimProb} with $\rho$ fixed. In the following subsection, Section \ref{subsec:Implementation}, we detail our implementation of the penalty method for computing \eqref{eq:estimator} more broadly. 

\subsection{Majorize-minimize algorithm for computing $\tilde\mbB_\rho$}\label{subsec:MMAlg}
To compute \eqref{penalty_optimProb}, we appeal to the majorize-minimize principle. Following \citet{xu2017generalized}, given current ($k$th) iterate $\mbB^k$, we first majorize both distance penalties using a variation of 
$$\mathcal{M}_{\mathcal{C}_r}(\mbB \mid \mbB^k) = \|\mbB - \mathbf{P}_{\mathcal{C}_r}(\mbB^k)\|_F^2.$$
The function $\mathcal{M}_{\mathcal{C}_r}(\cdot \mid \mbB^k)$ majorizes ${\rm dist}(\cdot, \mathcal{C}_r)^2$ at $\mbB^k$ because 
\begin{equation} \label{eq:MM_conds}
{\rm dist}(\mbB^k, \mathcal{C}_r)^2 = \mathcal{M}_{\mathcal{C}_r}(\mbB^k \mid \mbB^k) \quad \text{ and } \quad   \mathcal{M}_{\mathcal{C}_r}(\mbB \mid \mbB^k) \geq {\rm dist}(\mbB, \mathcal{C}_r)^2
\end{equation}
for all $\mbB \in \mathbb{R}^{p \times J}.$ 
Moreover, $\nabla \mathcal{M}_{\mathcal{C}_r}(\mbB \mid \mbB^k) = 2\{\mbB - \mathbf{P}_{\mathcal{C}_r}(\mbB^k)\}$ when $\mathbf{P}_{\mathcal{C}_r}(\mbB^k)$ is single valued. An analogous majorizer can be constructed for ${\rm dist}(\cdot, \mathcal{A}_s)^2$.
Because both $\mathcal{C}_r$ and $\mathcal{A}_s$ are closed, projections onto each are single valued except on a set of matrices with Lebesgue measure zero \citep[Proposition 6]{keys2019proximal}. In addition, both projections can be computed efficiently. Letting $\boldsymbol{L}^k_r \in \mathbb{R}^{p \times r}$ and $\boldsymbol{R}^k_r \in \mathbb{R}^{J \times r}$ denote the leading $r$ left and right singular vectors of $\mbB^k$, respectively, and letting $\boldsymbol{D}^k_r \in \mathbb{R}^{r \times r}$ be a diagonal matrix with $\mbB^k$'s largest $r$ singular values along its diagonal, 
$\mathbf{P}_{\mathcal{C}_r}(\mbB^k) = {\boldsymbol{L}^k_{r}} {\boldsymbol{D}^k_{r}} {\boldsymbol{R}^k_{r}}^\top.$
Similarly, the projection of $\mbB^k$ onto $\mathcal{A}_s$ simply requires setting the rows with $(s+1)$th through $p$th largest Euclidean norms to zero. That is, $\mathbf{P}_{\mathcal{A}_s}(\mbB^k) = \mathbb{A}_s(\mbB^k) \mbB^k$ where $\mathbb{A}_s(\mbB^k) \in \mathbb{R}^{p \times p}$ is a diagonal matrix with $(l,l)$th element equal to one if $\|\mbB^k_{l,\cdot}\|_2$ is one of the $s$ largest for all $l\in [p]$ and zero otherwise. 

Letting $\mathcal{F}_\rho$ denote the objective function from \eqref{penalty_optimProb}, we can majorize $\mathcal{F}_\rho$ at $\mbB^k$ using that
\begin{equation}\label{eq:majoriziation}
\mathcal{F}_\rho(\boldsymbol{B}) \leq - \boldsymbol{\mathcal{L}}(\mbB) + \mu \|\mbB\|_F^2 + \frac{\rho}{2} \|\mbB - \mathbf{P}_{\mathcal{C}_r}(\mbB^k)\|_F^2 + \frac{\rho}{2}\|\mbB - \mathbf{P}_{\mathcal{A}_s}(\mbB^k)\|_F^2
\end{equation}
for all $\mbB \in \mathbb{R}^{p \times J}$. Thus if $\boldsymbol{B}^{k+1}$ minimizes the right hand side of \eqref{eq:majoriziation}, i.e., is defined as
\begin{equation}\label{eq:RidgeProblem}
\argmin_{\mbB \in \mathbb{R}^{p \times J}}  \left\{ - \boldsymbol{\mathcal{L}}(\mbB) + \mu \|\mbB\|_F^2  + \frac{\rho}{2} \|\mbB - \mathbf{P}_{\mathcal{C}_r}(\mbB^k)\|_F^2 + \frac{\rho}{2}\|\mbB - \mathbf{P}_{\mathcal{A}_s}(\mbB^k)\|_F^2 \right\},
\end{equation}
 then we would be ensured that $\mathcal{F}_\rho(\boldsymbol{B}^{k+1}) \leq \mathcal{F}_\rho(\boldsymbol{B}^{k})$ by \eqref{eq:MM_conds}. However, computing \eqref{eq:RidgeProblem} is itself a challenging optimization problem. Although \eqref{eq:RidgeProblem} can be solved column-by-column of $\mbB$, each column-wise update would require its own iterative algorithm to compute. 
To avoid solving this problem at each iteration, we can instead employ a further (approximate) majorization of $-\boldsymbol{\mathcal{L}}(\mbB)$, which can be minimized efficiently. Let 
$$\ell_{(j)}(\mbb_{(j)}) =\sum_{i=1}^{n_{(j)}}\delta_{(j)i}\bigg[\mbx_{(j)i}^\top\mbb_{(j)} - \log \Big\{ \sum_{k \in \mathcal{R}_{(j)i}} {\rm exp}\left( \mbx_{(j)k}^\top\mbb_{(j)} \right) \Big\} \bigg]$$
denote the $j$th population's contribution to the partial log-likelihood for $j \in [J]$. 
Following \citet{simon2011regularization}, we use a quadratic approximation of $-\ell_{(j)}(\mbb_{(j)}).$ Let $\mbeta_{(j)}^k = \mbX_{(j)}\mbb_{(j)}^k$ where $\mbX_{(j)} = (\mbx_{(j)1}, \dots,\mbx_{(j)n_{(j)}})^\top\in \mathbb{R}^{n_{(j)} \times p}$. For square matrix $\mbA$, let ${\rm Diag}(\mbA)$ be a matrix with $(j,k)$th entry equal to $\mbA_{j,k}$ if $j = k$ and zero otherwise. Then, letting $f_{(j)}$ denote the function $\mbeta_{(j)} \mapsto -\ell_{(j)}(\mbb_{(j)})$, and letting $\nabla f_{(j)}$ and $\nabla^2 f_{(j)}$ denote the gradient and Hessian of $f_{(j)}$, we have that 
\begin{align*} -\ell_{(j)}(\mbb_{(j)}) \approx \hspace{5pt} & -\ell_{(j)}(\mbb_{(j)}^k) + \left(\mbX_{(j)}\mbb_{(j)} - \mbeta_{(j)}^k\right)^\top\nabla f_{(j)}(\mbeta_{(j)}^k) \\
 & + \frac{1}{2}\left(\mbX_{(j)}\mbb_{(j)} - \mbeta_{(j)}^k\right)^\top{\rm Diag}\left[ \nabla^2 f_{(j)}(\mbeta_{(j)}^k)\right] \left(\mbX_{(j)}\mbb_{(j)} - \mbeta_{(j)}^k\right).
 \end{align*}
Hence, letting 
\begin{align*}
 g_{(j)}(\mbb_{(j)} \mid \mbb_{(j)}^k) &= -\ell_{(j)}(\mbb_{(j)}^k) + \left(\mbX_{(j)}\mbb_{(j)} - \mbeta_{(j)}^k\right)^\top\nabla f_{(j)}(\mbeta_{(j)}^k)  \\
 & \quad + \frac{1}{2}\left(\mbX_{(j)}\mbb_{(j)} - \mbeta_{(j)}^k\right)^\top{\rm Diag}\left[ \nabla^2 f_{(j)}(\mbeta_{(j)}^k)\right] \left(\mbX_{(j)}\mbb_{(j)} - \mbeta_{(j)}^k\right) + \frac{\mu}{2}\mbb_{(j)}^\top\mbb_{(j)},
 \end{align*}
 and letting $[\mathbf{P}_{\mathcal{C}_r}(\mbB^k)]_{\cdot,j}$ denote the  
 $j$th column of the matrix $\mathbf{P}_{\mathcal{C}_r}(\mbB^k)$,
we define
\begin{equation}\label{eq:MM_objective} \mbb_{(j)}^{k+1} = \argmin_{\mbb_{(j)} \in \mathbb{R}^p} \left\{ g_{(j)}(\mbb_{(j)} \mid \mbb_{(j)}^k) + \frac{\rho}{2}\|\mbb_{(j)} - [\mathbf{P}_{\mathcal{C}_r}(\mbB^k)]_{\cdot,j}\|_2^2  + \frac{\rho}{2}\|\mbb_{(j)} - [\mathbf{P}_{\mathcal{A}_s}(\mbB^k)]_{\cdot,j} \|_2^2 \right\}
\end{equation}
for $j \in [J]$,
and define $\mbB^{k+1} = (\mbb_{(1)}^{k+1}, \dots, \mbb_{(J)}^{k+1})$. Minimizing this particular approximation of $\mathcal{F}_\rho$ is efficient because each $\mbb_{(j)}^{k+1}$ can be computed in parallel and in closed-form. 
Letting $$\mbZ(\mbeta_{(j)}^k) = \mbX_{(j)}\mbb_{(j)}^k -{\rm Diag}\{ \nabla^2 f_{(j)}(\mbeta_{(j)}^k)\}^{-1}\nabla f_{(j)}(\mbeta_{(j)}^k), \quad \mbW(\mbeta_{(j)}^k) = {\rm Diag}\{ \nabla^2 f_{(j)}(\mbeta_{(j)}^k)\},$$ and $\lambda = 2\rho + \mu$, we have that 
\begin{equation} \label{eq:b_update}
\mbb_{(j)}^{k+1} = \left\{\mbX_{(j)}^\top\mbW(\mbeta_{(j)}^k)\mbX_{(j)} + \lambda I_p\right\}^{-1} \left\{ \mbX_{(j)}^\top\mbW(\mbeta_{(j)}^k) \mbZ(\eta_{(j)}^k) + \rho [\mathbf{P}_{\mathcal{C}_r}(\mbB^k) + \mathbf{P}_{\mathcal{A}_s}(\mbB^k)]_{\cdot,j}\right\}.
\end{equation}
Computing \eqref{eq:b_update} may be time consuming when $p > n_{(j)}$. In this case, we may use the Woodbury identity, i.e.,
$\{\mbX_{(j)}^\top\mbW(\mbeta_{(j)}^k)\mbX_{(j)} +  \lambda I_p\}^{-1} = \lambda^{-2}[ \lambda I_p - \mbX_{(j)}^\top\{\mbW(\mbeta_{(j)}^k)^{-1} + \lambda^{-1}\mbX_{(j)}\mbX_{(j)}^\top\}^{-1}\mbX_{(j)}],$
so that computing $\mbb_{(j)}^{k+1}$ requires inverting only an $n_{(j)} \times n_{(j)}$ matrix.

We summarize the algorithm for computing $\tilde{\mbB}_\rho$ (with $\rho$ fixed) in Algorithm \ref{alg:MM} and embed this within the penalty method for computing \eqref{eq:estimator} in Algorithm \ref{alg:penalty}.  Note that our approach is not strictly adhering to the majorize-minimize principle. The objective function constructed from summing the terms in \eqref{eq:MM_objective} over all $j$ is not, strictly speaking, a majorizing function of \eqref{penalty_optimProb}.
To ensure that $\mbB^{k+1} = (\mbb_{(1)}^{k+1}, \dots, \mbb_{(J)}^{k+1})$ yields a decrement of the original objective function $\mathcal{F}_\rho$, one could replace each ${\rm Diag}\{ \nabla^2 f_{(j)}(\mbeta_{(j)}^k)\}$ with $\phi I_{n_{(j)}}$ for some sufficiently large positive constant $\phi$. In practice, we found that replacing ${\rm Diag}\{ \nabla^2 f_{(j)}(\mbeta_{(j)}^k)\}$ is not necessary. Updates based on \eqref{eq:MM_objective} led to a decrease in the objective function in every scenario we considered, and led to substantially faster convergence than any other approach we tried. However, this modified version of our algorithm is amenable to convergence analysis. 
\begin{prop}\label{prop:conv}
Let $\mbB^{k+1}= (\mbb_{(1)}^{k+1}, \dots, \mbb_{(J)}^{k+1})$ be defined as in \eqref{eq:MM_objective} with 
$\phi I_{n_{(j)}}$ replacing 
${\rm Diag}\{\nabla^2 f_{(j)}(\mbeta_{(j)}^{k})\}$ for some $\phi > 0$ sufficiently large and fixed. Then,  $\mathcal{F}_{\rho}(\mbB^{k+1}) \leq \mathcal{F}_{\rho}(\mbB^{k})$ for $k = 1, 2, 3, \dots$,  $\lim_{k\to \infty} \|\mbB^{k+1} - \mbB^{k}\|_F = 0$, and if each limit point of $\{\mbB^{k}\}_{k=1}^\infty$ is isolated, then the iterates converge to a critical point of $\mathcal{F}_{\rho}$.
\end{prop}
Proposition \ref{prop:conv} -- which is a direct application of Proposition 9 of \citet{keys2019proximal} -- follows, in part, from the majorize-minimize principle and the strong convexity of $g_{(j)}(\cdot \mid \mbb^{k}_{(j)})$.

\begin{algorithm}[t]\caption{Majorize-minimize algorithm for \eqref{penalty_optimProb}}\label{alg:MM}
\textbf{Inputs:} $(\mu, r, s, k^{\rm max}, {\mbB}^{\rm init},\rho) \in (0, \infty) \times [\min(p,J)] \times [p] \times \mathbb{N} \times \mathbb{R}^{p \times J} \times (0, \infty)$
\vspace{-10pt}
\begin{enumerate}
	\item  Set $\mbB^0 = {\mbB}^{\rm init}$ and $k = 0$
\item Compute $\breve{\mbB} = \mathbf{P}_{\mathcal{C}_r}(\mbB^k) + \mathbf{P}_{\mathcal{A}_s}(\mbB^k)$
\item  For $j \in [J]$ in parallel
\begin{enumerate}
	\item Update $\mbeta_{(j)}^k = \mbX_{(j)}\mbb_{(j)}^k$
\item Update $\mbb_{(j)}^{k+1} = \left\{\mbX_{(j)}^\top\mbW(\mbeta_{(j)}^k)\mbX_{(j)} +  (2\rho + \mu) I_p\right\}^{-1} \left\{ \mbX_{(j)}^\top\mbW(\mbeta_{(j)}^k) \mbZ(\mbeta_{(j)}^k) + \rho \breve\mbb_{(j)}\right\}$
\end{enumerate}
\item If the objective function value has not converged and $k \leq k^{\rm max}$, set $k = k + 1$ and return to (2); otherwise, terminate
\end{enumerate}
\end{algorithm}

\subsection{Implementation details}\label{subsec:Implementation}
To use the penalty method, we must apply Algorithm \ref{alg:MM} to repeatedly solve \eqref{penalty_optimProb} for an increasing sequence of penalty parameters $\rho$. Starting with a relatively small initial value of $\rho > 0$, we compute $\tilde{\mbB}_\rho$ using Algorithm \ref{alg:MM}. We then increase $\rho$ by a fixed multiplicative factor, say 1.2, and compute $\tilde{\mbB}_{1.2\rho}$ after initializing the algorithm at $\tilde{\mbB}_\rho$. We repeat this procedure until for some sufficiently large $\rho$ and corresponding $\tilde{\mbB}_{\rho}$, we have both
${\rm dist}(\tilde{\mbB}_{\rho}, \mathcal{C}_r)^2 < \epsilon$ and ${\rm dist}(\tilde{\mbB}_{\rho}, \mathcal{A}_s)^2 < \epsilon$
for some convergence tolerance $\epsilon > 0$. We call the limit of this sequence $\widehat\mbB$, which can be thought of as \eqref{eq:estimator} for a fixed $r,s,$ and $\mu$.

Following \citet{keys2019proximal}, we find that it is not necessary to solve \eqref{penalty_optimProb} exactly for each penalty parameter $\rho$. Instead, we can run Algorithm \ref{alg:MM} for a fixed number of iterations (e.g., we set $k^{\rm max} = 10$ in our implementation) to approximate \eqref{penalty_optimProb} before incrementing $\rho$. Often, 10 iterations was enough for convergence for even moderately sized $\rho$, e.g., $\rho > (1.2)^5 \rho_0$, where $\rho_0$ is the initial penalty parameter value. The number of allowed iterations, $\rho_0$, and by what factor $\rho$ is increased at each iteration of the penalty method can be diagnosed by the user, but we found our default implementation to work well across a range of scenarios. Software implementing this method is available for download from GitHub \citep{IntegrativeCox}.

In addition, we found that using ``warm-starting'' could also improve estimation accuracy and convergence speed. Suppose that we consider models defined by $(r,s) \in \left\{r_1, \dots, r_L\right\} \times \left\{s_1, \dots, s_M\right\}$ where $r_1 > r_2 > \dots > r_L$. Treating $s = s_i$ as fixed, we start by computing $\widehat{\mbB}$ with $r = r_1$ after initializing at the matrix of zeros. Then, for computing $\widehat{\mbB}$ with $r = r_k$, we initialize the algorithm at the solution for $r = r_{k-1}$ for $k \in \{2, \dots, L\}$.  We repeat this separately for each candidate $s$. 

\begin{algorithm}[t]\caption{Penalty method for computing \eqref{eq:estimator}}\label{alg:penalty}
\textbf{Inputs:} $(\mu, r, s, k^{\rm max}, \delta_{\rm incr}, \rho_0, \epsilon) \in (0, \infty) \times [\min(p,J)] \times [p] \times \mathbb{N} \times (1, \infty) \times (0, \infty) \times (0, \infty)$
\vspace{-10pt}
\begin{enumerate}
\item Initialize $\tilde\mbB \in \mathbb{R}^{p \times J}$ and set $\rho_{\rm old} = \rho_0$
\item Compute $\tilde{\mbB}_{\rho_{\rm new}}$, the final iterate of Algorithm 1 with inputs $(\mu, r, s, k^{\rm max}, \tilde\mbB, \rho_{\rm old})$
\item Set $\tilde{\mbB}_{\rho_{\rm old}} = \tilde{\mbB}_{\rho_{\rm new}}$ and set $\rho_{\rm new} = \delta_{\rm incr}  \rho_{\rm old}$
\item Compute $\tilde{\mbB}_{\rho_{\rm new}}$, the final iterate of Algorithm 1 with inputs $(\mu, r, s, k^{\rm max}, \tilde\mbB_{\rho_{\rm old}}, \rho_{\rm new})$
\item If $\max\{{\rm dist}(\tilde{\mbB}_{\rho_{\rm new}}, \mathcal{C}_r)^2, {\rm dist}(\tilde{\mbB}_{\rho_{\rm new}}, \mathcal{A}_s)^2\} < \epsilon$, terminate; otherwise, set $\rho_{\rm old} = \rho_{\rm new}$ then return to (3)
\end{enumerate}
\end{algorithm}

\section{Asymptotic properties}\label{sec:asymp}
Next, we establish the asymptotic distribution of a version of our estimator. We focus on quantifying the efficiency gained when using the rank constraint by studying \eqref{eq:estimator} with $s = p$ and $\mu = 0$. We treat $p$ as fixed throughout this section. To simplify notation, let $\widehat{\mbB}_r$ denote the rank $r$ constrained version of \eqref{eq:estimator}, i.e., define $\widehat{\mbB}_r = \argmax_{\mbB \in \mathcal{C}_r \cap \mathcal{D}_M} \boldsymbol{\mathcal{L}}(\mbB)$ where $\mathcal{D}_M = \{\boldsymbol{B}\in \mathbb{R}^{p \times J}: \|\mbB\|_F \leq M\}$ for large constant $M$. The constraint that $\widehat{\mbB}_r \in \mathcal{D}_M$  serves to regularize the estimator in a manner similar to the ridge penalty. However, we omit notation indicating dependence on $M$ since, for the remainder of this section, we assume $M$ is taken to be a arbitrarily large constant such that $\|\mbB_*\|_F \ll M < \infty$. 

In the following, we establish asymptotic normality,  derive the asymptotic variance, and verify that the covariance of each column of the standard maximum partial likelihood estimator minus that of our estimator is positive semidefinite. Together, these results provide a theoretical justification of the efficiency gains observed in our simulation studies in Section \ref{sec:SimulationStudies}. Throughout, let $\mbA^{+}$ be the Moore-Penrose pseudoinverse of a matrix $\mbA$, and let $\mbA_1 \otimes \mbA_2$ be the Kronecker product of matrices $\mbA_1$ and $\mbA_2.$ Let ${\rm BlockDiag}[\{\mbA_k\}_{k=1}^K]$ be the block diagonal matrix with $k$th diagonal block $\mbA_k$ for $k \in [K]$, and let ${\rm vec}(\mbA)$ be the vector formed by stacking the columns of the matrix $\mbA$. Finally, we use the notation $\mbA_1 \succeq \mbA_2$ to indicate that $\mbA_1 - \mbA_2$ is positive semidefinite.

Our asymptotic results will require a number of assumptions.  First, we assume that ${\rm rank}(\mbB_*) = r_*$ with $r_* \leq \min\{p,J\}$, and assume that $r = r_*$ (i.e., the rank of \eqref{eq:estimator} is correctly specified). Similarly, letting $n = \sum_{j=1}^J n_{(j)}$, we assume that for all $j \in [J]$, the ratio $n_{(j)}/n \to \kappa_{(j)}$ for some constant $\kappa_{(j)} > 0$ as $n \to \infty$. The remainder of our assumptions apply to each population separately: these are the standard assumptions needed for asymptotic normality under the Cox proportional hazards model with right censoring, e.g., see Chapter 6 of \citet{MR1915446} or \citet{hjort2011asymptotics}. Notably, we do not require anything about the relationships between populations other than the low-rankness of $\mbB_*$.  For the sake of space, we formally state and discuss these assumptions, (A1)--(A7),  in Web Appendix D.  

Before stating our main result, we first note that when $r = \min\{p,J\}$, $\widehat{\boldsymbol{B}}_r$ is equivalent to $J$ separate maximum partial likelihood estimators: one for each column of $\boldsymbol{B}_*$. We denote this estimator $\bar{\boldsymbol{B}}$ and define $\bar{\mbb}_{(j)} \in \mathbb{R}^p$ as the $j$th column of $\bar{\mbB}$ for $j \in [J].$ Applying standard asymptotic results for the Cox proportional hazards model, we know that as $n \to \infty$ under our assumptions, $\sqrt{n_{(j)}}(\bar{\boldsymbol{b}}_{(j)} - \boldsymbol{b}_{*(j)}) \xrightarrow{d} {\rm N}_p(0, \mbD_{*(j)}^{-1})$ for $j \in [J]$, where the exact form of $\mbD_{*(j)}$ is given in Lemma E.2 of  Web Appendix E. As we will show, the asymptotic distribution of $\widehat{\mbB}_{r_*}$ will depend on the $\mbD_{*(j)}$. However, unlike the unconstrained maximum partial likelihood estimator $\bar{\mbB}$, all $\mbD_{*(j)}$ affect the asymptotic covariance of each column of $\widehat{\mbB}_{r_*}$. The following result, proved in Web Appendix D, establishes the asymptotic distribution of the rank-constrained maximum partial likelihood estimator. 
\begin{theorem}\label{thm:consistency}
Let $\boldsymbol{U} \in \mathbb{R}^{p \times r_*}$ and $\boldsymbol{V} \in \mathbb{R}^{J \times r_*}$ be any pair of rank $r_*$ matrices such that $\mbB_* = \U\V^\top$. Define the matrices
$\mbT = [\V \otimes I_p, I_J \otimes \U]$ and $\mbD_* = {\rm BlockDiag}\left[\{\kappa_{(j)} \mbD_{*(j)}\}_{j=1}^J\right].$
Then under assumptions (A1)--(A7),
$$ \sqrt{n} \{ {\rm vec}(\widehat{\boldsymbol{B}}_{r_*} - \boldsymbol{B}_*)\} \xrightarrow{d} {\rm N}_{Jp}\left( 0, \mbT(\mbT^\top\mbD_* \mbT)^+ \mbT^\top\right).$$
\end{theorem}
Before discussing the result, we comment briefly on the proof. There are two fundamental differences between our proof and those of existing asymptotics for reduced-rank regression estimators. Firstly, in contrast to the classical reduced-rank regression asymptotics \citep{stoica1996maximum,anderson1999asymptotic}, we do not have closed-form expressions for estimates of a particular $\U$ and $\V$. Secondly, we are not dealing with a quadratic objective function, so many other standard techniques could not be applied. Instead, we prove Theorem \ref{thm:consistency} by first defining a particular identifiable decomposition of $\mbB_*$, showing that we estimate the components of this particular decomposition consistently, using this to establish the asymptotic normality of a function of these components, and finally showing that these results hold for any decomposition of $\mbB_*$. 

To better understand the implication of Theorem \ref{thm:consistency}, we express the asymptotic covariance in terms of any pair $(\U, \V)$ where $\mbB_* = \U \V^\top$. Defining $\tilde{\mbD}_{*(j)} = \kappa_{(j)} \mbD_{*(j)}$ for all $j \in [J]$, $\mbP_{\mbA,\mbB} = \mbA(\mbA^\top  \mbB \mbA)^{-1}\mbA^\top\mbB$, and $\mbP_{\mbA, \mbB}^\perp = I - \mbP_{\mbA, \mbB}$ for any projection matrix $\mbP_{\mbA, \mbB}$, we have that asymptotic covariance from Theorem \ref{thm:consistency} can be expressed 
\begin{align*}
 \mbT&(\mbT^\top\mbD_* \mbT)^+ \mbT^\top   = {\rm BlockDiag}\left[ \{\mbP_{\U, \tilde\mbD_{*(j)}}\tilde\mbD_{*(j)}^{-1}\}_{j=1}^J \right]  \\
& ~~~~~~~~+ \left(\begin{array}{c}
        \V_{1,\cdot}^\top \otimes \mbP^\perp_{\U, \tilde\mbD_{*(1)}} \\
        \V_{2,\cdot}^\top \otimes \mbP^\perp_{\U, \tilde\mbD_{*(2)}} \\
        \vdots\\
        \V_{J,\cdot}^\top \otimes \mbP^\perp_{\U, \tilde\mbD_{*(J)}}
        \end{array}\right) \left(\sum_{j=1}^J \V_{j,\cdot}\V_{j,\cdot}^\top \otimes \tilde\mbD_{*(j)} \mbP^\perp_{\U, \tilde\mbD_{*(j)}} \right)^{+}\left(\begin{array}{c}
        \V_{1,\cdot}^\top \otimes \mbP^\perp_{\U, \tilde\mbD_{*(1)}}\\
        \V_{2,\cdot}^\top \otimes \mbP^\perp_{\U, \tilde\mbD_{*(2)}} \\
        \vdots\\
         \V_{J,\cdot}^\top \otimes \mbP^\perp_{\U, \tilde\mbD_{*(J)}}
        \end{array}\right)^\top 
\end{align*}
 where $\V_{j,\cdot} \in \mathbb{R}^r$ denotes the $j$th row of $\V.$
That is, the asymptotic covariance consists of the sum of two matrices: a block diagonal matrix whose components depend only on the $\tilde\mbD_{*(j)}$ and the column space of $\U$, and a matrix which depends on the $\tilde\mbD_{*(j)}$ and the column spaces of $\U$ and $\V$. The covariance between columns of $\sqrt{n}(\widehat{\mbB}_{r_*} - \mbB_*)$ comes only from the latter matrix. For example, the covariance between the $l$th and $m$th columns is
$$ \left(\V_{l,\cdot}^\top \otimes \mbP^\perp_{\U, \widetilde\mbD_{*(l)}}\right) \left( \sum_{j=1}^J \V_{j,\cdot} \V_{j, \cdot}^\top \otimes\widetilde\mbD_{*(j)} \mbP_{\U, \widetilde\mbD_{*(j)}}^\perp   \right)^{+}\left(\V_{m, \cdot}^\top \otimes \mbP^\perp_{\U, \widetilde\mbD_{*(m)}}\right)^\top.$$
Note that the asymptotic covariance of the maximum partial likelihood estimator is rank deficient. It can be checked that ${\rm rank} [{\rm avar} \{ {\rm vec}(\widehat{\mbB}_{r_*} - \mbB_*)\}] \leq (J + p - r_*)r_*,$ where $(J + p - r_*)r_*$ is the number of identifiable parameters in $\mbB_*$ when ${\rm rank}(\mbB_*) = r_*$ and ${\rm avar}$ denotes the asymptotic covariance.

Finally, we can use the result of Theorem \ref{thm:consistency} to verify that the rank restriction leads to an efficiency gain over separate maximum partial likelihood estimators.
\begin{theorem}\label{corol:efficiency}
Under the conditions of Theorem \ref{thm:consistency}, 
$$ {\rm avar}\{\sqrt{n}(\bar{\boldsymbol{b}}_{(j)} - \mbb_{*(j)})\} \succeq {\rm avar}\{\sqrt{n}(\widehat{\boldsymbol{b}}_{(j)r_*} - \mbb_{*(j)})\}~~ \text{for all } ~ j \in[J],$$
where $\widehat{\boldsymbol{b}}_{(j)r_*}$ is the $j$th column of $\widehat{\mbB}_{r_*}$.
\end{theorem}
Theorem \ref{corol:efficiency} reveals that each column of the rank-constrained estimator has covariance no greater than the unconstrained maximum partial likelihood estimator in the sense that their difference is negative semidefinite.  
The result of Theorem \ref{corol:efficiency} suggests that the improvements in estimation accuracy, which we observe empirically in Section \ref{sec:SimulationStudies}, are the result of lower variance coming from the rank restriction.

\section{Simulation studies}\label{sec:SimulationStudies}
\subsection{Data generating model}\label{sec:DataGeneratingModel}
We compare our method to various competitors under the assumption in \eqref{eq:RR}. Specifically, for one hundred independent replications, we generate survival times under the Cox proportional hazards models for $J=12$ distinct populations. In each setting, we generate $n_{(1)} = n_{(4)} = n_{(7)} = n_{(10)} = 100$, $n_{(2)} = n_{(5)} = n_{(8)} = n_{(11)} = 200$, and $n_{(3)} = n_{(6)} = n_{(9)} = n_{(12)} = 300$
independent survival times for each population. For each subject in our dataset, we first generate predictors $\mbx_{(j)i}$ as a realization of ${\rm N}_p(0, \boldsymbol{\Sigma})$ where $\boldsymbol{\Sigma}_{l,m} = 0.7^{|l-m|}$ for $(l,m) \in [p] \times [p].$ Given $\mbx_{(j)i}$, we then generate the true survival time according to the Cox proportional hazards model with Gompertz baseline hazard using $T_{(j)i} = \log\{ 1 -  (\alpha/\zeta_{(j)}) \log(u) {\rm exp}(-\mbx_{(j)i}^\top\mbb_{*(j)})\}/\alpha$ for $i \in [n_{(j)}]$ and $j \in [J]$, 
where $u \sim {\rm Uniform}(0,1)$ independently for all $(j)i$ combinations. We set $\alpha = {\pi}/({600\sqrt{6}})$ and $\zeta_{(j)} = \alpha \hspace{2pt}\exp\{-0.5772 - \alpha \nu_{(j)} \}$ where $\nu_{(j)} = 2000 + 10(j - 1)$ for $j \in [J]$. These parameter values are chosen so that the generated survival times mimicked those in the TCPA data analyzed in Section \ref{sec:DataAnalysis}. Under this data generating model, the baseline hazards are distinct across populations. See \citet{bender2005generating} for more details about this data generating model. Given $t_{(j)1}, \dots, t_{(j)n_{(j)}}$ for $j\in [J]$, we generate censoring times $c_{(j)i}$ as realizations of an exponential random variable with mean $q_{\xi}(\{t_{(j)i}\}_{i=1}^{n_{(j)}}),$
where $q_{\xi}$ denotes the $\xi$th quantile of its argument. We allow $\xi$ to vary across populations. If $n_{(j)} < 300$, we use $\xi = \tau$, whereas if $n_{(j)} \geq 300$, we use $\xi = \tau + 0.20$ with $\tau$ varying across simulation settings.   

Given censoring times, we set $y_{(j)i} = \min\{c_{(j)i}, t_{(j)i}\}$ and $\delta_{(j)i} = \mathbf{1}(y_{(j)i} = t_{(j)i})$. We generate $\mbB_* = \boldsymbol{UV}^\top \in \mathbb{R}^{p \times J}$ where $\U \in \mathbb{R}^{p \times r_*}$ has $20$ rows randomly selected to be nonzero with each nonzero entry independent and uniformly distributed on $[-\sqrt{8}/r_*, -\sqrt{2}/r_*] \cup [\sqrt{2}/r_*, \sqrt{8}/r_*]$. The matrix $\V \in \mathbb{R}^{r_* \times J}$ is a randomly generated semi-orthogonal matrix.

In our simulation studies, we consider: (1) $r_* \in [6]$ with $p = 250$ and $\tau = 0.35$; (2) $\tau \in \left\{0.25, 0.35, 0.45, 0.55, 0.65\right\}$ with $p = 250$ and $r_* = 3$; and (3) $p \in \left\{100, 200, 300, 400, 500\right\}$ with $r_* = 3$ and $\tau = 0.35$. In each replication, we also construct validation and testing sets of size $150$ and $1000,$ respectively, for each $j \in [J]$. We consider three performance metrics: (1) ${\rm tr}\{(\widehat\mbB - \mbB_*)^\top \boldsymbol{\Sigma}(\widehat\mbB - \mbB_*)\}$, i.e., model error; (2) concordance (C-index) between linear predictors and observed survival times; and (3) Brier score evaluated at the median observed survival time. Concordance and Brier scores are averaged over the $J$ populations. See Web Appendix C for further descriptions of these metrics. 
 
\subsection{Competing methods}
We consider various competing methods, some of which can exploit the low-rankness assumption in \eqref{eq:RR}. Many of the competitors we consider estimate $\mbB_*$ column-by-column. The $j$th column of these estimators can be expressed
\begin{equation} \label{eq:sepEst}
\argmax_{\mbb_{(j)} \in \mathbb{R}^p}\left\{ n_{(j)}^{-1}\ell_{(j)}(\mbb_{(j)}) -  \lambda_{(j)} \|\mbb_{(j)}\|_q^q\right\},
\end{equation}
where $\lambda_{(j)} \in (0, \infty)$ is a user specified tuning parameter for $j \in [J]$ and $q \in \left\{1, 2\right\}$. 
Based on \eqref{eq:sepEst}, we consider the following alternative estimators:
\texttt{Sep-Ridge}, \eqref{eq:sepEst} with $q = 2$ and each $\lambda_{(j)}$ chosen to minimize the deviance on the $j$th population validation set (i.e., separate ridge regressions used for each population); \texttt{Sep-Lasso}, \eqref{eq:sepEst} with $q = 1$ and $\lambda_{(j)}$ chosen to minimize the deviance on the $j$th population validation set;  \texttt{Proj-Sep-Ridge} (resp. \texttt{Proj-Sep-Lasso}), the nearest rank-$r$ approximation to separate ridge-penalized (resp. lasso-penalized) Cox models. To compute \texttt{Proj-Sep-Ridge}, we first obtain \eqref{eq:sepEst} with $q = 2$ and each $\lambda_{(j)}$ chosen to minimize the deviance on the $j$th population validation set, then find the nearest rank $r$ approximation (in squared Frobenius norm) to the estimate.  The rank parameter, $r$, is also chosen to minimize the deviance on the validation set. The latter estimators are inspired by the ``lazy'' sparse reduced-rank regression estimator from \citet[Section 3.4.5]{qian2020large}.

We also consider \texttt{Convex-Approx}, the nuclear norm and group lasso penalized estimator of $\mbB_*$ described in Web Appendix A.1.  Note that both \texttt{Proj-} estimators and \texttt{Convex-Approx} exploit the assumption of low-rankness in $\mbB_*$. Finally, we use our method with $\rho_0 = 50$ (where $\rho_0$ is the initializing value of $\rho$ for the penalty method) and $\mu = 0.1$ fixed across every setting. Tuning parameters $s$ and $r$ are chosen to minimize the validation set deviance. We use \texttt{LR-Cox} to denote \eqref{eq:estimator}. 

\subsection{Results}
\begin{figure}[t!]
\centering
\includegraphics[width=\textwidth]{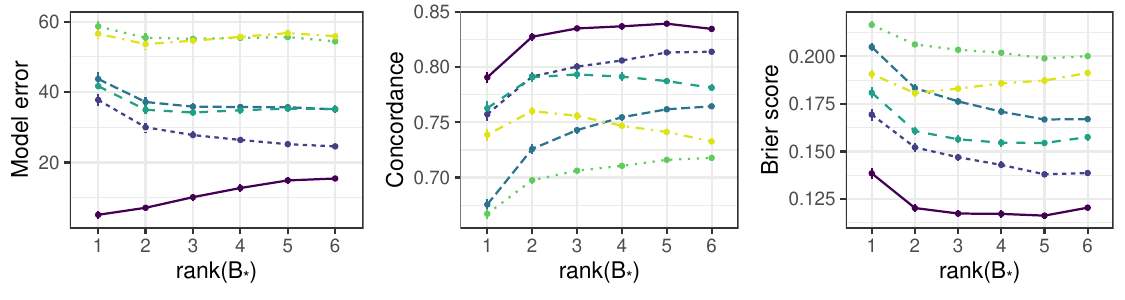}
\includegraphics[width=\textwidth]{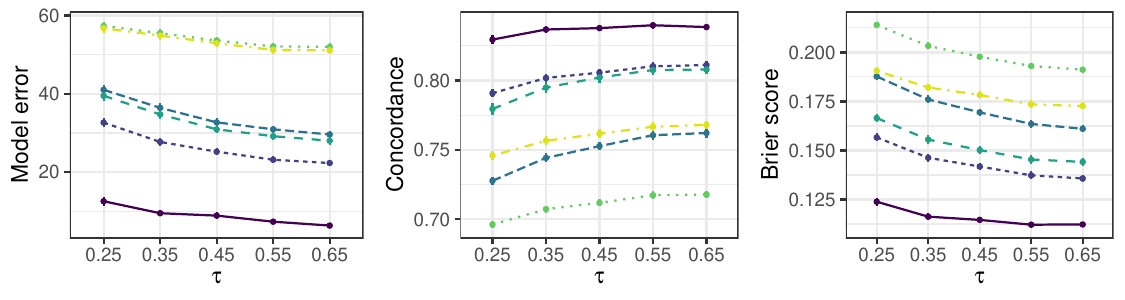}
\includegraphics[width=\textwidth]{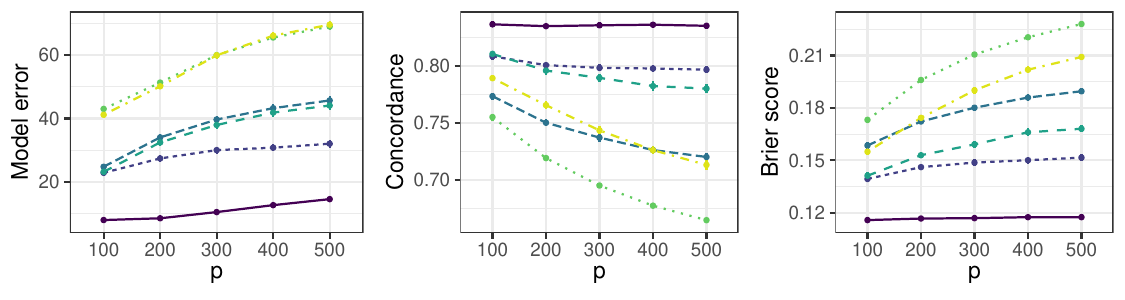}
\includegraphics[width=14cm]{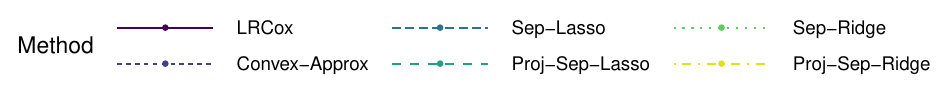}
\caption{Averages (plus and minus two standard errors) for each of the six methods over 100 independent replications under the models described in Section \ref{sec:DataGeneratingModel} with (top row)  $(p, \tau) = (250, 0.35)$, (middle row) $(p, r_*)  = (250, 3)$, and (bottom row) $(r_*, \tau) = (3, 0.35)$. This figure appears in color in the electronic version of this article.}\label{fig:SimulationResults}
\end{figure}

We display results in Figure \ref{fig:SimulationResults}. In the first row, we display the performance of the estimators as the rank of the true regression coefficient matrix increases with $\tau$ and $p$ held fixed at $0.35$ and $250$, respectively. In the three settings we consider, we see that in terms of model error, concordance, and Brier score, our method performs the best among all the competitors considered. As the rank increases, the difference in model error between our method and \texttt{Convex-Approx} begin to decrease. In terms of model error, only the performance of \texttt{LR-Cox} degrades as the rank increases. In contrast, when considering both concordance and Brier score,  performance of every method (including the proposed \texttt{LR-Cox}) improves as the rank of $\mbB_*$ increases. This can be explained by the fact that under our data generating model, as the rank increases, the signal strength increases. Nevertheless, in each setting, our method outperforms all competitors.

In the second row of Figure \ref{fig:SimulationResults}, we display results as $\tau$, the censoring parameter, increases with $p$ and the rank of $\mbB_*$ held fixed at $250$ and 3, respectively. We see the performance of all methods improve in all metrics as the proportion of censoring decreases, but our method's superior performance remains constant across the considered $\tau$. Notably, \texttt{Proj-Sep-Lasso} only slightly outperforms \texttt{Convex-Approx} in terms of concordance, although \texttt{Convex-Approx} is better than \texttt{Proj-Sep-Lasso} in terms of Brier score and model error. 

Finally, in the bottom row of Figure \ref{fig:SimulationResults}, we display results as $p$ varies with $\tau$ and the rank of $\mbB_*$ held fixed at 0.35 and 3, respectively. As expected, as $p$ grows, keeping the number of truly important variables fixed at 20, each method's performance tends to degrade. However, the performance of \texttt{LR-Cox} seems to be less affected by the change in $p$ than does the convex approximation \texttt{Convex-Approx} or any of the competing estimators. For example, in terms of Brier score, our method tends to perform similarly for all considered values of $p$, whereas  performance of all other methods degrades as $p$ approaches 500. 

Interestingly, the difference between \texttt{Convex-Approx} and \texttt{Proj-Sep-Lasso} is relatively small in all three settings. For example, \texttt{Proj-Sep-Lasso} outperforms \texttt{Convex-Approx} in terms of concordance when the rank of $\mbB_*$ is less than three. This lends further evidence to the notion that \texttt{Convex-Approx} leads to overly biased estimates. 

\subsection{Additional simulation studies}
In the Web Appendix, we provide additional simulation study results. In Web Appendix F.1, we compare \eqref{eq:estimator} to an alternative estimator which uses a sample size-weighted version of $\boldsymbol{\mathcal{L}}$. In Web Appendix F.2, we assess the sensitivity of \eqref{eq:estimator} to the choice of rank. Unsurprisingly, we see that overspecification of the rank has only a slight effect relative to underspecification. In Web Appendix F.3, we consider comprehensive simulation studies under three alternative data generating models. Specifically, we consider settings where some factors are not shared by subsets of the populations and settings where some predictors are relevant for only a subset of the populations. Finally, in Web Appendix F.4, we present mean squared estimation error results under the data generating models from Section \ref{sec:DataGeneratingModel}.

\begin{figure}[t]
\centering
\includegraphics[width=15cm]{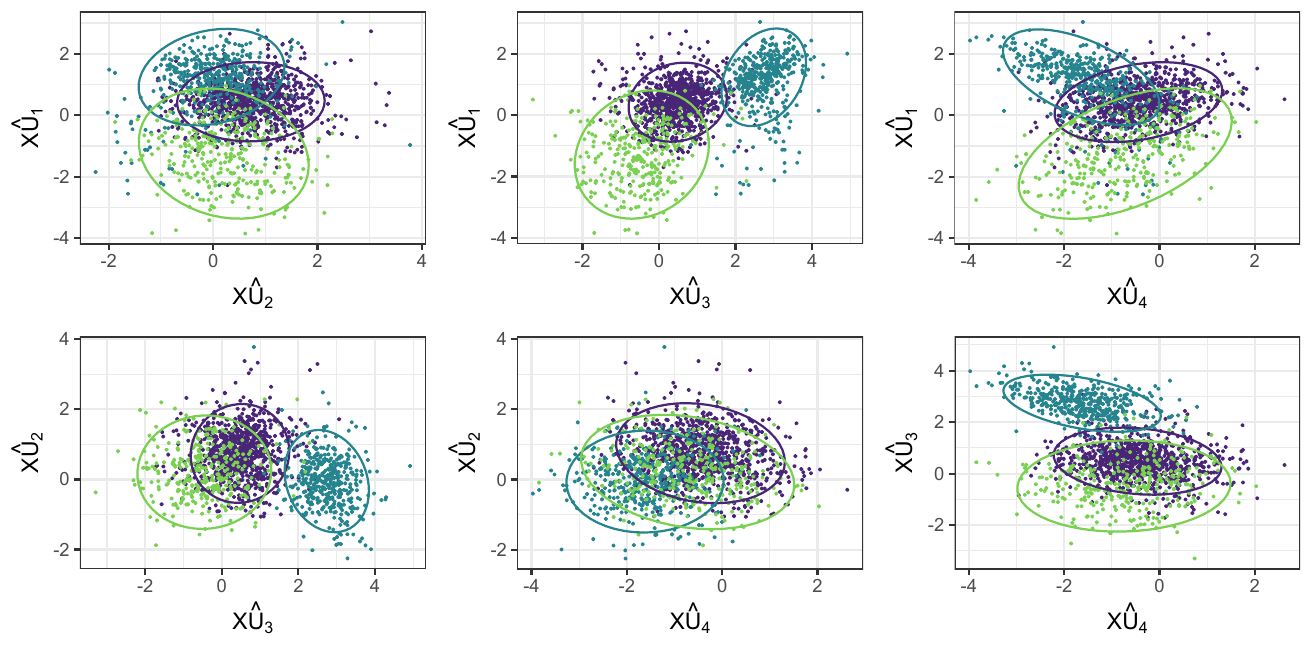}
\caption{Pairwise estimated factors for three cancer types: BRCA (purple), LGG (blue), and LUSC (green). Normal ellipses, included for improved visualization, were computed using \texttt{stat\_ellipse} in \texttt{ggplot2}. This figure appears in color in the electronic version of this article: the mentioned colors refer to that version.}\label{fig:factors}
\end{figure}

\begin{figure}[h]
\centering
\includegraphics[width=12cm]{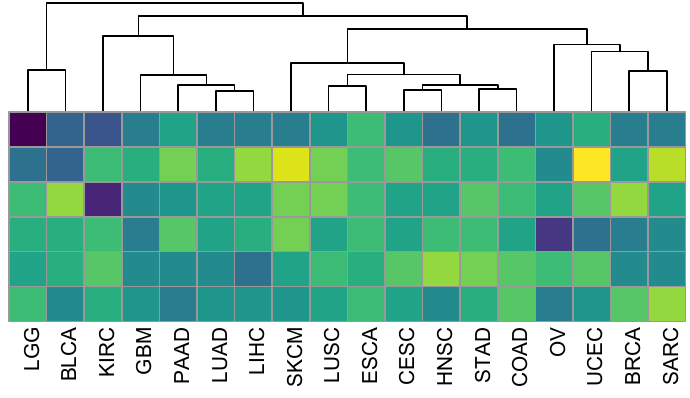}\hspace{10pt}\includegraphics[width=1cm]{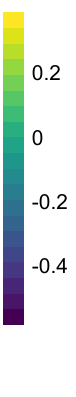}
\caption{A heatmap of $\widehat{\boldsymbol{D}}\widehat{\boldsymbol{R}}^\top$ based on the fitted model described in Section \ref{sec:Results}. Cancer types were sorted by hierarchical clustering. This figure appears in color in the electronic version of this article.}\label{fig:DV}
\end{figure}

\section{Pan-cancer survival analysis with protein expression}\label{sec:DataAnalysis}
\subsection{Data processing}\label{sec:DataProcessing}
Finally, we perform the pan-cancer integrative survival analysis that motivated our proposed method. The data we analyze is from The Cancer Proteome Atlas (TCPA), detailed in \citet{li2013tcpa}. These data consist of clinical information and protein expression measurements from reverse-phase protein arrays for patients with many distinct types of cancer. In our analysis, we use data from all cancer types which had at least 30 patients with recorded failure times and 30 with censored failure times. Kaplan-Meier survival curves for each of the 18 cancer types are displayed in Web Figure 1. The normalized expression (level 4 from Pan-Can 32 from the TCPA database\nocite{TCPA}) of the $p = 210$ proteins with no missing expression in the cancer types we analyzed were used as the predictors in our analysis.

\subsection{Results}\label{sec:Results}
We use five-fold cross-validation to select $r$ and $s$ for model fitting. For each $k \in [5]$, we fit the model using all the data outside the $k$th fold with $r \in [10]$, $s \in \left\{10, 12, 14, \dots, 40\right\}$, $\mu = 50$, and $\rho_0 = 50$. In Web Figure 2, we display a heatmap of the cross-validated linear predictor scores (see equation (13) of the Web Appendix). Models with rank less than four all performed poorly, as did models with rank greater than seven and $s \geq 26$. The minimum overall cross-validated linear predictor score is achieved by $s = 20$ and $r = 6$. 

We refit the model to the entire dataset using $s = 20$ and $r=6$. This model assumes there are $r = 6$ linear combinations of the proteins and each linear combination consists of the same $s = 20$ proteins. Compared to the full model with $pJ = 3780$ coefficients to be estimated, there are $(J + s -r)r = 192$ parameters to be estimated in our model. See Web Appendix I for an explanation of the number of parameters in reduced-rank models. 

First, we display the estimated factors to show how they separate distinct types of cancer. Specifically, taking the singular value decomposition $\widehat\mbB =\widehat{\U}\widehat{\boldsymbol{D}}\widehat{\R}^\top$ where $\widehat{\U} = (\hat{\mbu}_1, \dots, \hat{\mbu}_6) \in \mathbb{R}^{p \times 6}$, $\mbx^\top \hat{\mbu}_{k}$ can be interpreted as the $k$th estimated factor for a subject with protein expression $\mbx \in \mathbb{R}^p$. One can think of these factors as the underlying low-dimensional predictors, and can in turn think of $\widehat{\V}^\top = \widehat{\boldsymbol{D}}\widehat{\boldsymbol{R}}^\top$ as the regression coefficient matrix for these low-dimensional predictors. In Figure \ref{fig:factors}, we display all pairwise factors for three cancer types: breast cancer (BRCA, purple), low grade glioma (LGG, blue), and lung squamous cell carcinoma (LUSC, green). We see that the first and third factors nearly entirely separate the three cancer types. In addition, we see that BRCA and LUSC are entirely separated in three of the six plots. 

In Figure \ref{fig:DV}, we also display the matrix $\widehat{\boldsymbol{D}}\widehat{\boldsymbol{R}}^\top$. Since $\widehat{\boldsymbol{D}}\widehat{\boldsymbol{R}}^\top$ can be interpreted as the matrix of regression coefficients corresponding to the low-dimensional factors, we can compare coefficients across cancer types. For example, KIRC has a negative coefficient for the third factor, whereas BRCA has a positive coefficient. Similarly, BLCA has a negative coefficient for the second factor, whereas UCEC has a relatively large positive coefficient.

The proteins selected by our method can be found in Table \ref{table:proteins}. 
Many of the identified proteins are known to play a role in cancer biology. For example, c-Kit, a tyrosine kinase receptor, is often found in higher amounts on the surface of cancer cells.  Recent studies have shown that c-Kit is expressed in aggressive cancers, on circulating tumor cells, and in recurrent and resistant tumors \citep{foster2018cd117}. Similarly, there is evidence that FASN, which encodes fatty acid synthase, is a metabolic oncogene that plays a central role in tumor progression and survival \citep{flavin2010fatty}. In addition, Annexin A1 is known to inhibit innate immune cells and promote T-cell activation \citep{perretti2009annexin}, and Paxillin is believed to play a role in tumorigenesis and metastasis \citep{deakin2012diverse}.

To assess whether our method yields better fitted models than competitors on rare cancer types, we also perform a leave-one-out cross-validation analysis. In terms of both concordance and linear predictor scores, our method performs better than \texttt{Sep-Lasso}, \texttt{Sep-Ridge}, and separate elastic net estimators on four of the five cancer types we considered. In the one cancer type where a competitor performed better, none of the methods had concordance higher than 0.5, which corresponds to randomly guessing the linear predictor ordering. Additional details can be found in Web Appendix G.

\subsection{External validation of factors}
To further investigate whether our method identifies a useful set of pan-cancer proteomic factors for survival, we use our estimated factors to model survival in other cancer types. Specifically, we focus on four rare cancer types from TCPA which were excluded from our analysis in Section \ref{sec:Results}. These are cancer types that have at least $50$ subjects, and both $15$ failure and censoring times: kidney renal papillary cell carcinoma (KIRP), thyroid carcinoma (THCA), rectum adenocarcinoma (READ), and mesothelioma (MESO), which have sample sizes $208, 372, 130,$ and $61$,  respectively. 

\begin{table}
\caption{Proteins with nonzero coefficient estimates using the full dataset with $(r,s) = (6, 20)$ chosen by five-fold cross-validation minimizing the cross-validated linear predictor score.}\label{table:proteins}
\begin{tabular}{lllll}
\toprule
 x4E-BP1\_pT37\_T46 & c-Kit  & Caspase-7\_cleavedD198  & Caveolin-1\\  
   Gab2  & HSP70 & IGFBP2  & MAPK\_pT202\_Y204  \\
   PAI-1 & Paxillin  & FASN  & MYH11   \\
   TFRC & EPPK1 & Acetyl-a-Tubulin-Lys40  & Annexin-1 \\ 
   EGFR & NF-kB-p65\_pS536 & NDRG1\_pT346 & p16Ink4a\\
   \bottomrule
  \end{tabular}
\end{table}
\bigskip

\begin{table}
\caption{Average concordance (and standard errors) for each of the four considered methods across 1000 independent replications in each of the four datasets. Bolded cells are those with largest median among the four methods.}\label{tab:external_validation}
\centering
\begin{tabular}{c||cccc}     
\toprule 
Cancer & \texttt{LR-Cox-DR} & \texttt{Sep-Ridge} & \texttt{Sep-Lasso} & \texttt{Sep-En}\\
\hline
KIRP & 0.575 (0.007) & \textbf{0.672} (0.007) & 0.464  (0.003) & 0.497 (0.006) \\ 
  THCA & \textbf{0.545}  (0.008) & 0.528  (0.009) & 0.456  (0.004) & 0.460  (0.007) \\ 
  READ & \textbf{0.529} (0.009) & 0.524  (0.009) & 0.490  (0.005) & 0.534  (0.008) \\ 
  MESO & \textbf{0.618} (0.006) & 0.551 (0.006) & 0.526  (0.004) & 0.544  (0.005) \\ 
  \bottomrule 
  \end{tabular}
\end{table}

For 1000 independent replications, we randomly split each dataset into a training set (90\%) and testing set (10\%). Recall that in these data, predictors consist of $p = 210$ proteins. We fit a Cox proportional hazards model to the training data using three methods: \texttt{Sep-Lasso}, \texttt{Sep-Ridge}, and a version of \eqref{eq:sepEst} with elastic net penalty (\texttt{Sep-En}). For each method, tuning parameters are chosen by five-fold cross-validation on the training set. We obtain the estimated linear predictors on the testing set based on the fitted model using the tuning parameter which minimized partial likelihood deviance (the default in \texttt{glmnet}).

The fourth method we considered, \texttt{LR-Cox-DR}, relies on our dimension-reduced fitted model from Section \ref{sec:Results}. Specifically, letting $\widehat{\U}$ be the left singular vectors of the estimate of $\mbB_*$ from before, we first set $\tilde{\mbx}_i = \widehat{\U}^\top \mbx_i \in \mathbb{R}^6$ (where $\mbx_i$ is the $i$th training subject's predictors) and fit a standard Cox proportional hazards model with $\tilde{\mbx}_i$ as predictors. We do this separately for each of the four cancer types.  Then, we transform testing set predictors using the same $\widehat{\U}$ and obtain the estimated linear predictor based on the fitted Cox model. It is important to emphasize that the estimate $\widehat{\U}$ came from datasets entirely separate from those we consider here. Specifically, $\widehat{\U}$ corresponds to the estimated factors from the 18 cancer types analyzed in Section \ref{sec:Results} which did not include KIRP, THCA, READ, or MESO.

In each replication and cancer type, we measure the concordance between the estimated linear predictor and the true survival outcomes (with appropriate adjustments for censoring). We report averages in Table \ref{tab:external_validation}. We see that our method, denoted \texttt{LR-Cox-DR} in Table \ref{tab:external_validation}, performs as well or better than competitors in three of the four external datasets. Ridge regression performs best in one dataset (KIRP), and in another (READ), \texttt{LR-Cox-DR}, \texttt{Sep-En}, and \texttt{Sep-Ridge} were not significantly different. When considering only methods which perform variable selection, \texttt{LR-Cox-DR} performs as well or better than \texttt{Sep-Lasso} and \texttt{Sep-En} in all four datasets. Together, these results suggest that our estimated factor weights may generalize to other cancer types reasonably well.

\section{Discussion}
There are two directions we plan to explore in future research. First, our estimator relies on a computational approach using distance-to-set penalties. Instead, one could compute \eqref{eq:estimator} using an iterative hard thresholding algorithm \citep{jain2014iterative}. This could be more efficient than our approach but would require iteratively projecting onto the intersection of the set of row-sparse and low-rank matrices, which is nontrivial.
Second, we plan to extend the proposed methodology to models which do not assume proportional hazards, e.g., the accelerated failure time (AFT) model. Recent advances in computation for semiparametric AFT models in high dimensions \citep{suder2021scalable} may be useful for such extensions.

\section*{Acknowledgements}
The authors thank two anonymous referees and the associate editor for their helpful comments. 
The authors also thank Karl Oskar Ekvall, Wei Sun, Adam J. Rothman, and Arun K. Kuchibhotla
for helpful conversations. A. J. Molstad's contributions were supported in part by a grant from the National Science Foundation (DMS-2113589).

\section*{Data availability statement}
The data that support the findings in this paper are openly available from The Cancer Proteome Atlas at https://tcpaportal.org/tcpa/ \citep{li2013tcpa,li2017explore}.

\bibliographystyle{biom} 
\bibliography{IntegrativeCox}

\section*{Supporting Information}
Web appendices, figures, and tables referenced in Sections 2, 4, 5, and 6 are available with
this paper at the Biometrics website on Wiley Online
Library. Code to reproduce all simulation results are  available for download at \url{github.com/ajmolstad/IntegrativeCox}. \vspace*{-8pt}
\vspace{-15pt}

\label{lastpage}

\end{document}